\documentclass[graybox]{svmult}

\usepackage{mathptmx}       % selects Times Roman as basic font
\usepackage{helvet}         % selects Helvetica as sans-serif font
\usepackage{courier}        % selects Courier as typewriter font
\usepackage{type1cm}        % activate if the above 3 fonts are
\usepackage{makeidx}         % allows index generation
\usepackage{graphicx}        % standard LaTeX graphics tool
\usepackage{multicol}        % used for the two-column index
\usepackage[bottom]{footmisc}% places footnotes at page bottom

\makeindex             % used for the subject index
\def\lsim{\hspace{0.3em}\raisebox{0.4ex}{$<$}\hspace{-0.75em}\raisebox{-.7ex}{$\sim$}\hspace{0.3em}}

\begin{document}

\title*{Vibrational Instability of Metal-Poor Low-Mass Main-Sequence Stars}
\author{Takafumi SONOI and Hiromoto SHIBAHASHI}
\institute{Takafumi SONOI \& Hiromoto SHIBAHASHI 
\at Department of Astronomy, University of Tokyo, 7-3-1 Hongo, Bunkyo-ku, 
Tokyo 113-0033, Japan  \email{sonoi@astron.s.u-tokyo.ac.jp} \& {shibahashi@astron.s.u-tokyo.ac.jp}}
%\and Hiromoto SHIBAHASHI \at Department of Astronomy, University of Tokyo, %7-3-1 Hongo, Bunkyo-ku, 
%Tokyo 113-0033, Japan \email{shibahashi@astron.s.u-tokyo.ac.jp}}
\maketitle
% Please use both starred abstract and non-starred abstract.

\abstract*{
We find that low-degree low-order g-modes become unstable in metal-poor low-mass stars due to the $\varepsilon$-mechanism of the pp-chain. 
Since the outer convection zone of these stars is limited only to the very outer layers, 
the uncertainty in the treatment of convection does not affect the result significantly. The decrease in metallicity leads to decrease in opacity and hence increase in
luminosity of a star. This makes the star compact and results in decrease in the density contrast, which is
favorable to the $\varepsilon$-mechanism instability.
We find also instability for high order g-modes of metal-poor low-mass stars by the convective blocking mechanism. Since the
effective temperature and the luminosity of metal-poor stars are significantly higher than those of Pop
I stars, the stars showing $\gamma$ Dor-type pulsation are substantially less
massive than in the case of Pop I stars. We demonstrate that those modes are unstable for about
$1\,M_\odot$ stars in the metal-poor case.
}

\abstract{
We find that low-degree low-order g-modes become unstable in metal-poor low-mass stars due to the $\varepsilon$-mechanism of the pp-chain. 
Since the outer convection zone of these stars is limited only to the very outer layers, 
the uncertainty in the treatment of convection does not affect the result significantly. The decrease in metallicity leads to decrease in opacity and hence increase in
luminosity of a star. This makes the star compact and results in decrease in the density contrast, which is
favorable to the $\varepsilon$-mechanism instability.
We find also instability for high order g-modes of metal-poor low-mass stars by the convective blocking mechanism. Since the
effective temperature and the luminosity of metal-poor stars are significantly higher than those of Pop
I stars, the stars showing $\gamma$ Dor-type pulsation are substantially less
massive than in the case of Pop I stars. We demonstrate that those modes are unstable for about
$1\,M_\odot$ stars in the metal-poor case.
}

\section{Introduction}
\label{sec:1}
The structure and evolution of stars born in metal-poor environments or when heavy elements were significantly deficient in the universe are considerably different than for stars born much later with higher heavy-element abundances. This is mainly because opacities and nuclear reactions are highly dependent on metallicity. Low metallicity leads to decreased opacity, and hence it makes luminosity of the star higher. On the other hand, the decrease in metallicity makes CNO-cycle energy generation less efficient. To maintain energy equilibrium in this situation, the stars with low metallicity need to be  compact compared with Pop I stars. As a result, the main sequence of metal deficient or metal poor stars on the HR diagram moves toward the bluer and higher luminosity side. 
The difference in structure must be conspicuous particularly for stars about $1.5\,M_{\odot}$, around which the main nuclear energy source changes between the CNO-cycle and the pp-chain and also the thickness of convective envelope drastically changes. Oscillation properties and stability of the stars are also expected to change significantly with metallicity. To confirm this, we perform a fully nonadiabatic analysis of metal-poor low-mass main-sequence stars.

\section{$\varepsilon$-mechanism for low-degree low-order g-modes}
\label{sec:2}

\begin{figure}[t]
  \begin{tabular}{cc}
    \begin{minipage}{0.5\hsize}
      \includegraphics[width=\hsize]{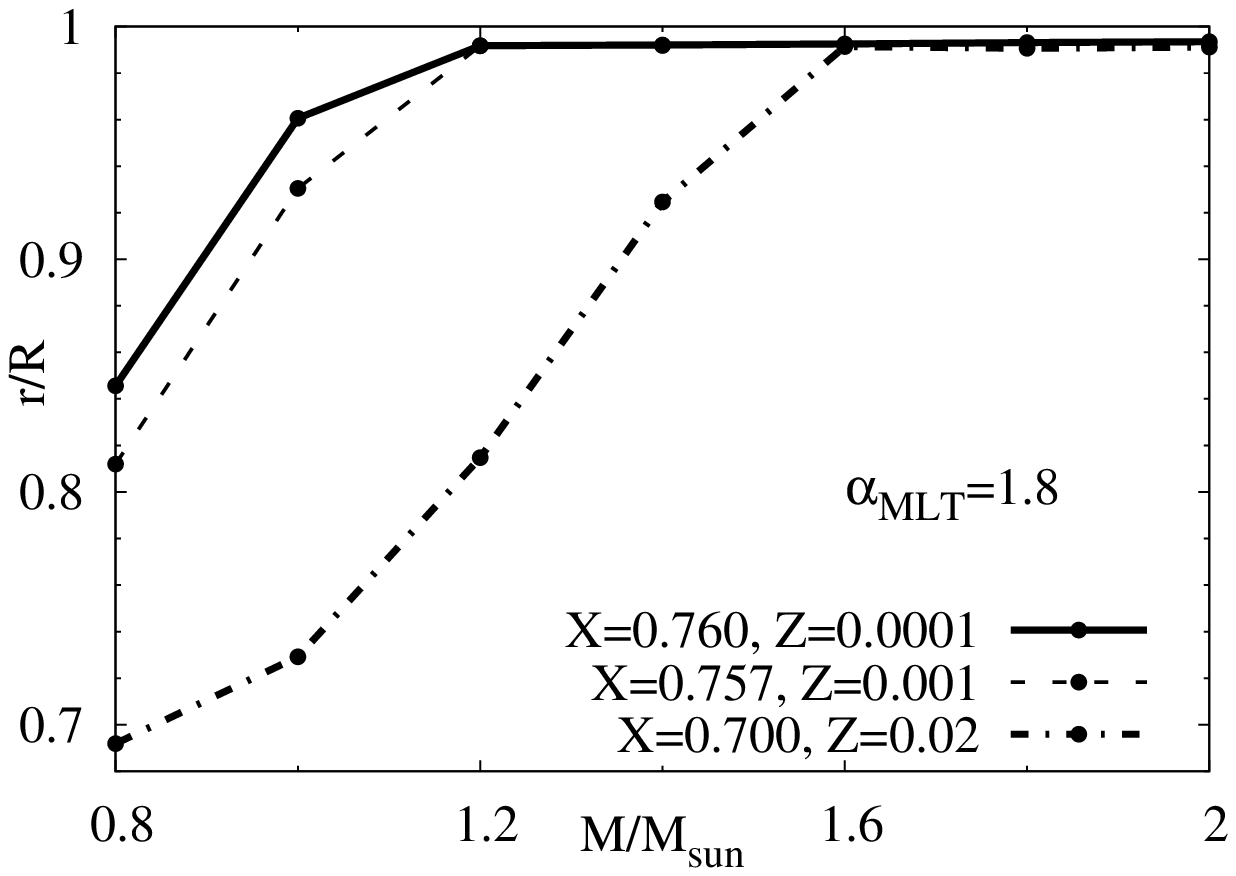}
    \end{minipage}
    \begin{minipage}{0.5\hsize}
      \includegraphics[width=\hsize]{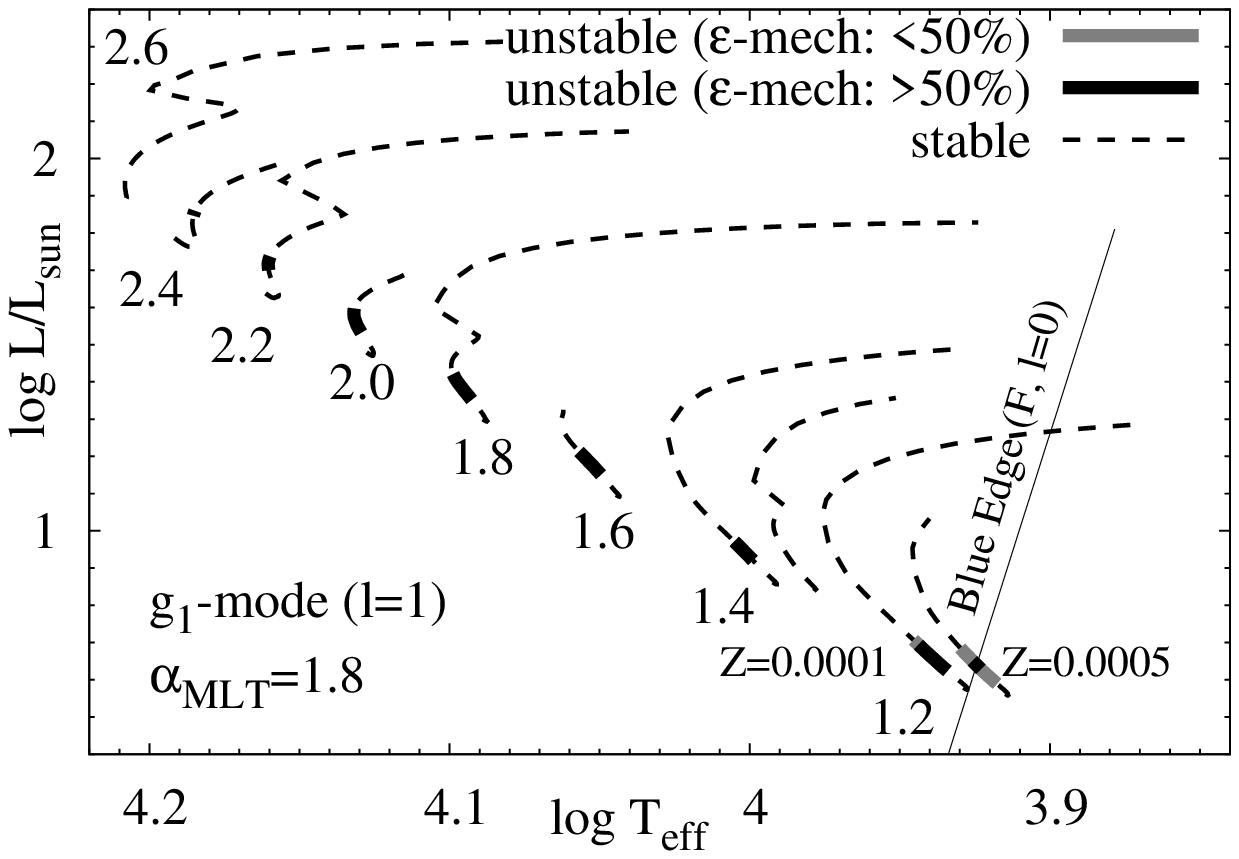}
    \end{minipage}
  \end{tabular}
  \caption{{\bf Left:} Location of base of convective envelope for ZAMS models with different compositions. {\bf Right:} Evolutionary tracks on HR diagram for $1.2-2.6\,M_{\odot}$ with $X=0.760,\; Z=0.0001$ and $1.2-1.4\,M_{\odot}$ with $X=0.759,\; Z=0.0005$. 
The {\it dashed} and {\it thick solid lines} indicate the evolutionary stages at which the dipole ($l=1$) g$_1$-mode is stable and unstable, respectively. The evolutionary stage in which the g$_1$-mode is unstable mainly due to the $\varepsilon$-mechanism is shown by {\it black thick lines}, while the $\kappa$-mechanism by {\it gray thick lines.}}
  \label{fig:1}       
\end{figure}

In the pp-chain, at the stellar center $^3$He is consumed by $^3$He($^3$He,2H)$^4$He or by $^3$He($^4$He,$\gamma$)$^7$Be immediately after it is generated. 
However, since the $^3$He reactions are highly sensitive to temperature, they do not efficiently occur in the outer part of the nuclear burning region. As a consequence, $^3$He accumulates in an off-centered shell. 
In such a situation, the high temperature dependence of the $^3$He reactions might excite g-modes having a large amplitude in the off-centered $^3$He shell. 
Such instability and the resultant material mixing was once proposed as a possible solution to the solar neutrino problem (\cite{Dilke1972}), and the stability of the sun was examined by several groups (\cite{Dziembowski1973, Chris1974, Boury1975, Shibahashi1975}). The presence of a convective envelope, which occupies the outer 20--30\% in the case of Pop I low-mass stars, however, has made it hard to reach a definite conclusion on the stability.
But the situation is different in metal-poor stars. As the metallicity decreases, the convective envelope becomes thinner and limited only to very outer layers (left panel of Fig. \ref{fig:1}). In such a case, we can obtain a definite conclusion on the instability without the uncertainty due to the convective envelope. Indeed, the present authors showed that Pop III stars are unstable against low-order, dipole g-modes due to the $\varepsilon$-mechanism of the pp-chain (\cite{Sonoi2011, Sonoi2012}). 
To extend our analyses to stars with $Z \ne 0$ but still having only very thin convective envelope, we examine the stability of stars with extremely low metallicity; $Z=0.0005$ and $Z=0.0001$ 
with the help of the stellar evolution code MESA (\cite{Paxton2011}) and a nonadiabatic pulsation code (\cite{Sonoi2012}).

The right panel of Fig. \ref{fig:1} shows the results of our nonadiabtic analysis of those stars against the dipole ($l=1$) g$_1$-mode. A wide range of models are found to become unstable due to the $\varepsilon$-mechanism. By comparing the stability analyses of stars with two different metallicities, we find that as the metallicity decreases the mass range of this instability extends toward higher mass. 
The reason for this is the tendency for the CNO-cycle to be replaced by the pp-chain for more massive stars grows as the metallicity and the CNO abundance decreases. 
The core of the metal-poor stars is convective at the ZAMS stage even with the pp-chain burning because of the high central temperature, and the gravity waves are evanescent there. But as the convective core shrinks and disappears with stellar evolution, the gravity waves can then propagate in the nuclear burning region. This is favorable for the $\varepsilon$-mechanism to work for instability. 

We find that the stars with $M \lsim 1.2M_{\odot}$ become vibrationally unstable in the case of $Z=0.0005$, and those with $M \lsim 2.4M_{\odot}$ become unstable in the case of $Z=0.0001$. The more massive stars keep the convective cores because of the dominant contribution of the CNO-cycle rather than the pp-chain. This is not favorable for instability, and low-degree low-order g-modes are not excited indeed in those stars. It should be remarked that in stars with $\log T_{\rm eff} \lsim 3.95$, corresponding to the gray thick line parts of the evolutionary tracks in the right panel of Fig. \ref{fig:1}, low-degree low-order g-modes are excited mainly by the $\kappa$-mechanism.    

\section{Low-degree high-order g-modes ($\gamma$ Doradus type oscillations)}
\label{sec:3}
\begin{figure}[t]
  \begin{tabular}{cc}
    \begin{minipage}{0.5\hsize}
      \includegraphics[width=\hsize]{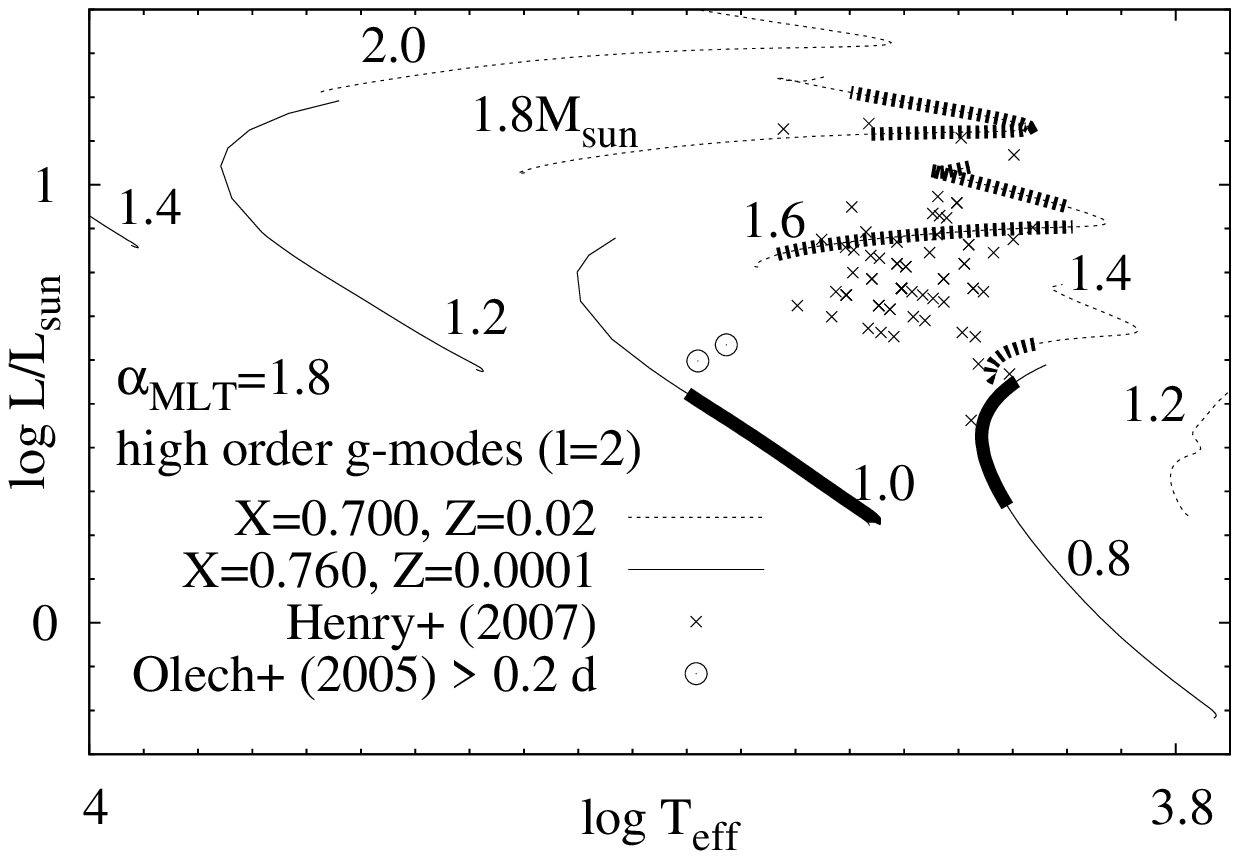}
    \end{minipage}
    \begin{minipage}{0.5\hsize}
      \includegraphics[width=\hsize]{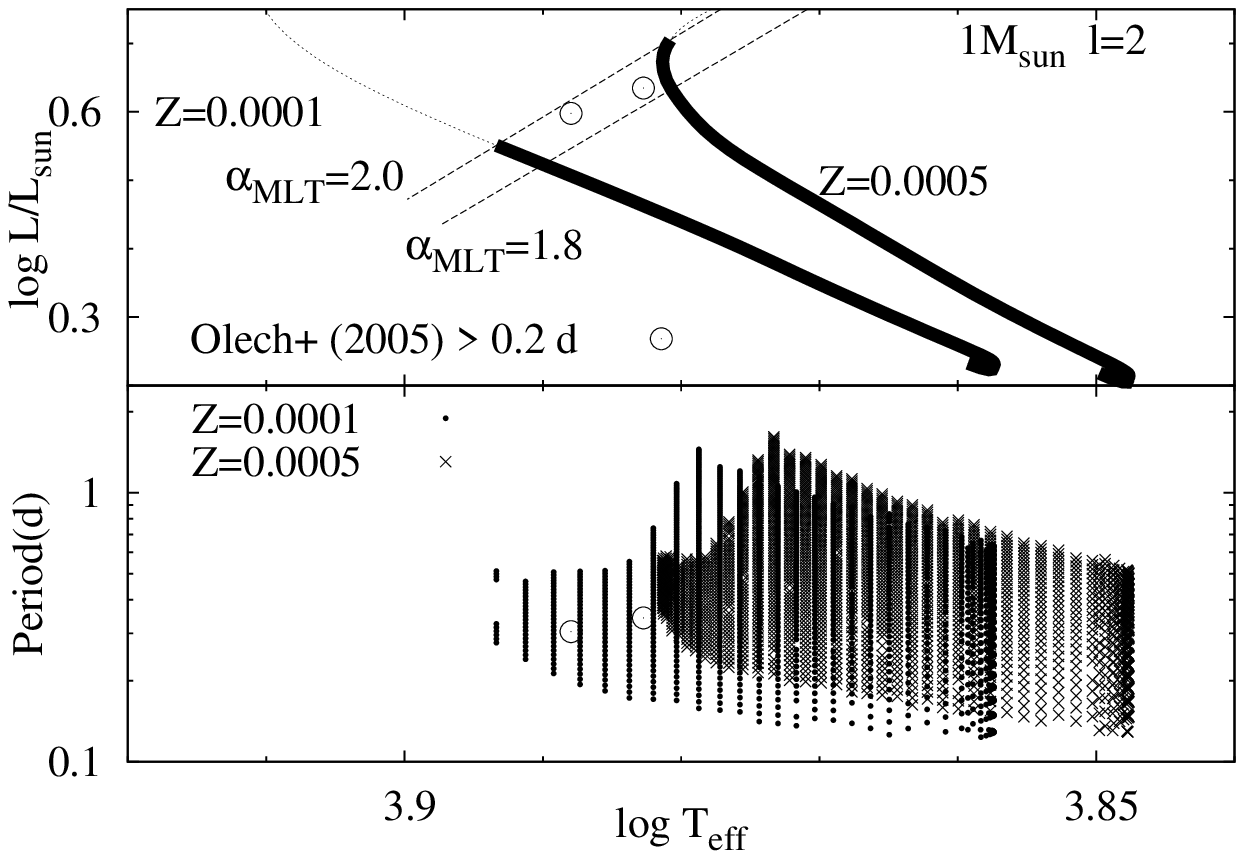}
    \end{minipage}
  \end{tabular}
  \caption{{\bf Left:} The {\it thick parts} of the evolutionary tracks correspond to unstable stages of high-order g-modes ($l=2$). The {\it crosses} are the $\gamma$ Dor stars reported by \cite{Henry2007}, while the {\it open circles} are stars, according to \cite{Olech2005}, having oscillation periods longer than 0.2 day. {\bf Top of Right:} Evolutionary tracks with $Z=0.0001$ and $0.0005$. 
The {\it thick parts} show the evolutionary stage in which high-order g-modes are excited. The {\it dashed lines} are the blue edges of the instability region of the high order g-modes ($l=2$) for $\alpha_{\rm MLT}=1.8$ and $2.0$, respectively. {\bf Bottom of right:} The {\it crosses} and {\it dots} denote periods of the unstable modes ($l=2$).}
  \label{fig:2}       
\end{figure}

Main-sequence stars near the cool end of the classical strip, called the $\gamma$ Dor stars, show low-degree high-order g-mode pulsations.
These stars have a thick convective envelope, and treatment of convection becomes an important issue in examining the stability analyses.   
Guzik et al. (2000) \cite{Guzik2000} noted the relatively long convective turn-over times at the base of the convective envelope of these stars. They claimed that a convective flux perturbation would be almost frozen-in and that excitation of the observed g-modes would be related with this characteristics and the fact that radiation becomes less responsible for energy transport from the base to the middle part of the convective envelope. 
Analyses with a time-dependent convection theory (\cite{Dupret2004}) demonstrated the above process.  Based on these ideas, we adopt here a ``frozen-in convection'' approximation and simply ignore the convective flux perturbation to examine the stability of the $\gamma$ Dor type oscillations of metal-poor stars. 
The instability region of high-order g-modes thus found in the metal-poor case is in the same temperature range as that in the Pop I case (Fig. \ref{fig:2}). But, since the evolutionary tracks move toward bluer and higher luminosity side as metallicity decreases, the instability appears for the less massive stars compared with the Pop I case.

Olech et al. (2005) \cite{Olech2005} found oscillations of blue stragglers in $\omega$ Cen, and detected many SX Phe stars, known as high amplitude $\delta$ Sct-like Pop II stars. They also detected oscillations having much lower frequencies than that of the radial fundamental mode. The corresponding stars are located around the blue edge of our theoretical instability region. Their periods correspond to the quadrupole ($l=2$) modes. Although the possibility exists that such low-frequency oscillations might be tidally induced, they might be the first examples corresponding to the modes investigated in the present analysis.

\begin{acknowledgement}
T. S. thanks the Hayakawa Satio Fund for its financial support.
\end{acknowledgement}

\end{document}